\documentclass[twoside]{article}

\usepackage{lipsum} 
\usepackage{aas_macros}

\usepackage[sc]{mathpazo} 
\usepackage[T1]{fontenc} 
\usepackage{microtype} 

\usepackage[hmarginratio=1:1,top=32mm,columnsep=20pt]{geometry} 
\usepackage{multicol} 
\usepackage[hang, small,labelfont=bf,up,textfont=it,up]{caption} 
\usepackage{booktabs} 
\usepackage{float} 
\usepackage{hyperref} 
\usepackage{natbib} 
\usepackage{graphicx}
\usepackage{caption} 
\usepackage{subcaption} 
\usepackage{lettrine} 
\usepackage{paralist} 
\usepackage{wasysym}
\usepackage{stmaryrd}

\usepackage{abstract} 

\usepackage{titlesec} 
\renewcommand\thesection{\Roman{section}} 
\renewcommand\thesubsection{\Roman{subsection}} 
\titleformat{\section}[block]{\large\scshape\centering}{\thesection.}{1em}{} 
\titleformat{\subsection}[block]{\large}{\thesubsection.}{1em}{} 

\usepackage{fancyhdr} 
\title{\vspace{-15mm}\fontsize{24pt}{10pt}\selectfont\textbf{Worlds in Migration}} 

\author{
\large
\textsc{Michael B. Lund$^1$}\\
\normalsize $^1$CalTech/IPAC-NExScI\\
\normalsize \href{mailto:mlund@ipac.caltech.edu}{mlund@ipac.caltech.edu} 
\vspace{-5mm}
}
\date{}

\begin{document}

\maketitle 

\thispagestyle{fancy} 


\begin{abstract}

\noindent In this paper we discuss an alternative track for migration that can explain the existence of Hot Jupiters observed in close orbits around their stars based on a novel interpretation of established work. We also discuss the population of sub-Earth rogue planets that would be created via this migration method, which would be on the order of 2 to 40 billion, many of which would still be present in the Galaxy and potentially detectable.

\end{abstract}


\begin{multicols}{2} 

\section{Introduction}
\lettrine[nindent=0em,lines=3]{I}n the current era of astronomy, discoveries of exoplanets have become frequent and almost ubiquitous. Though the existence of planets outside our solar system had been discussed as a concept (such as by \citet{Bruno1584} and \citet{Struve1952} to very different results), it was surprising when some of the first planets found differed greatly from our own Solar System's architecture, particularly as they tended to be Jupiter-like planets orbiting in very close orbits around their host stars \citep{Mayor1995}. 

This has naturally led to the challenge to understand how planetary formation processes can have such diversity as to produce not just our own solar system, but the myriad of planetary systems that have now been discovered. While there has been some suggestion that Hot Jupiters are able to form in situ close to their host stars \citep{Batygin2016}, there is also significant evidence that some Hot Jupiters must have migrated inward. Several discovered planets are known to be so close to their host stars that they have high rates of mass loss, such as GJ 3470b \citep{Bourrier2018}, WASP-12b \citep{Hebb2009, Lammer2009}, and KELT-9b \citep{Gaudi2017}. Multiple channels have been proposed for how Hot Jupiters migrate inward in their systems, including migration within the gas disk \citep{DAngelo2008} or through planet-planet interactions and the Kozai-Lidov mechanism \citep{Kozai1962, Lidov1962}. These pathways for Hot Jupiter formation have already been compared in great detail \citep{Dawson2018}.

In this paper we propose a renewed consideration of planetary migration as an extension of some ideas initially proposed in \citet{Velikovsky1950}, and which despite the criticism that they have received \citep{Goldsmith1977} may help explain migration as observed in Hot Jupiters when properly considered. We refer to this mechanism as Type V migration.

\section{Type V Migration} \label{TypeV}
While \citet{Velikovsky1950} does not present a pathway for Hot Jupiter migration explicitly, his work does outline a mechanism that can be applied to characterize Hot Jupiter migration. Velikovsky presented work on the sudden formation and migration of smaller terrestrial planets on short timescales, on the order of thousands of years. His work specifically focused on limited migration within our own solar system, explaining how electromagnetic forces could cause Venus to be ejected from within Jupiter's atmosphere before arriving on its current orbit. Velikovsky's hypothesis was in part based on prior work regarding the formation of new bodies directly from Jupiter (\citeauthor{Hesiod700}, 700BC).

We instead choose to focus on the influence that such actions would have on Jupiter, or a Jupiter-analogue, rather than on the smaller terrestrial body. The energy loss for the ejection of a single terrestrial object can be described as a function of the mass and velocity of the departing object:
\begin{equation}
    \label{eq:delE}
    \Delta E = \frac{1}{2}M_{\female}v^{2}
\end{equation}
Due to the scope of this work we fix M$_{\female}$ at 1 Venusian mass, or 0.815 M$_{\varoplus}$ \citep{Konopliv1999}. The velocity at time of ejection is a slightly more complex problem. We examine three different velocities, 100 km/s, 300 km/s, and 1000 km/s, the first two representing the velocities of most stars in the Milky Way, with the last being typical of a hypervelocity star \citep{Boubert2017, Hills1988}. Note that this velocity is combined with the escape velocities necessary for both escaping the mass of the Jovian planet and the host star (as 1 M$_{\varodot}$ in this work) and can be expressed using the equation for escape velocity:
\begin{equation}
\label{eq:Vesc}
    V_{esc} = \sqrt{\frac{2GM}{r}}
\end{equation}
Here the mass in question is of the primary object, and r represents the Jovian planetary radius or the orbital radius of the Jovian planet for escape velocities from the planet and from the star respectively. We treat the Jovian radius as $1 R_{J}$ for our purposes. These energy calculations neglect relativistic effects \citep{Einstein1905}.

By using the change in energy as calculated by \ref{eq:delE}, we can then determine how the orbital radius of the planet, in turn, changes by calculating the new orbital radius as a function of $\Delta \textrm{E, } \textrm{M}_{planet}\textrm{, } \textrm{M}_{\ast}$, and the current orbital radius, as given in Eq.\ref{eq:newr}.
\begin{equation}
    \label{eq:newr}
    r_{2} = \frac{1}{ \frac{1}{r_{1}} - \frac{2\Delta E}{GM_{planet}M{\ast}}  }
\end{equation}

\section{Simulations}\label{Simulations}
\begin{figure*}[!htb]
  \begin{center}
   \includegraphics[width=\textwidth]{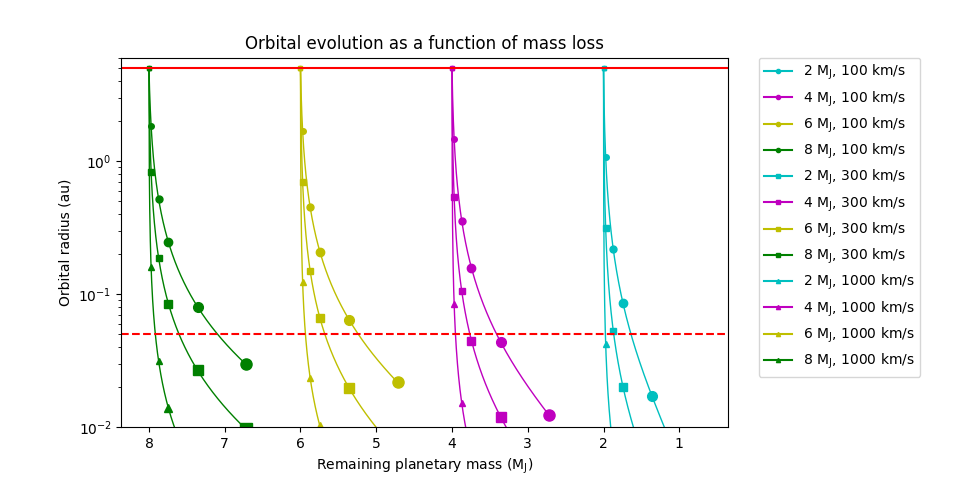}
  \end{center}
  \caption{The orbital radius of a Jovian planet as a function of remaining planetary mass for a range of initial planetary masses. For each initial planetary mass we examine three different tracks (as indicated by the symbols) that correspond to the resulting velocity of the ejected Venusian-mass bodies. The markers after the initial mass represent Jovian mass-loss corresponding to 10, 50, 100, 250, and 500 Venusian masses. Our starting point at the snow line is the red solid line at 5 AU, and the end point of 0.05 AU is the red dashed line.}
  \label{fig:migration}
\end{figure*}

In order to simulate the impact that Type V migration can have on a Jovian planet, we carry out a suite of planetary migration simulations. Our simulations use a range of Jovian planets between 2 and 8 $M_{J}$, all of which we place at a distance of 5 AU initially \citep{Jewitt2007}. We then eject 500 Venusian-mass planets with the three different final velocities mentioned in \S\ref{TypeV}, tracking how the planetary mass and orbital radius change, as shown in Figure~\ref{fig:migration}.

Our results show that it is quite possible that a Jovian-mass planet can lose sufficient mass and energy to migrate inward to the semi-major axis regime of Hot Jupiters while still maintaining sufficient mass to fit our observations. In the case of a 2M$_{J}$ planet, all tracks reach our final semi-major axis goal of 0.05 AU, with the slowest process requiring somewhat over 100 ejections. Alternatively, the hypervelocity ejections require fewer than 10 ejections from the 2M$_{J}$ planet to migrate to a 0.05 AU orbital radius. Even in the high-mass case of 8M$_{J}$, the 0.05 AU distance is reached for all tracks, although this does require more ejection events.

\section{Discussion}
The natural product of this Type V migration channel is a Galactic population of rogue planets. The existence of such planets has been discussed in several prior works, including some discussion of the dynamical histories required to create them \citep{Bear2000, Lissauer1987, Laughlin2000}. Type V migration necessitates that the Jovian planet remains in the system and only the terrestrial planet is actually ejected. Therefore, our described Type V migration mechanism is consistent with more recent microlensing observations that have found that while there are not many large rogue planets, there may be a large number of terrestrial-sized rogue exoplanets within the Milky Way \citep{Mroz2017}. These ejected planets fall well within the sensitivity range of the upcoming Wide-Field InfraRed Survey Telescope (WFIRST) microlensing mission \citep{Gehrels2015}, which will be sensitive to planets down to the size of Ganymede \citep{Penny2019}.

The estimated fraction of stars that host Hot Jupiters is around 1\% \citep{Wang2015}. With roughly 20 billion roughly sun-like stars in the galaxy \citep{Plait2013}, this translates to 200 million Hot Jupiters. Given the number of terrestrial planet ejections needed to generate this number of Hot Jupiters, our simulations suggest a population of 2-40 billion rogue Venusians in the Galaxy. We note, however, that if a significant fraction of these ejections involve hypervelocity planets, then the number still in the galaxy may be notably reduced as these bodies would be traveling in excess of galactic escape velocities, and so could become intergalactic planetary bodies \citep{BeastieBoys1998}.

While their fates have not been the primary focus of this work, these rogue Venusians (which have been speculated upon in previous works \citep{Lovecraft1939, Heinlein1951, Sterling1961, Adamski1977}) would also be prime candidates for Steppenwolf planets \citep{Abbot2011}, since they constitute a population of smaller terrestrial planets with thick insulating atmospheres that may be much more hospitable when not subject to the high irradiation that Venus receives.

\section{Summary}

In this paper we have proposed that the work of \citeauthor{Velikovsky1950} should be revisited in terms of how the ejection of small terrestrial planets from gas giants may provide an alternative migration track that can explain the observed population of Hot Jupiters. This would additionally indicate a population of rogue Venusian exoplanets numbering between 2 and 40 billion that may be detected by future microlensing missions such as WFIRST.

\section{Acknowledgements}
The author thanks Savannah R. Jacklin for her valuable feedback on this manuscript.

This research made use of Astropy,\footnote{http://www.astropy.org} a community-developed core Python package for Astronomy \citep{astropy:2013, astropy:2018}.

\bibliographystyle{apalike}
\bibliography{main}


\end{multicols}

\end{document}